\documentstyle[preprint,aps,floats]{revtex}
\begin{document}
\preprint{\tighten \vbox{
  \hbox{hep-th/0103020}}}
\draft
\title{On Critical Phenomena in a
Noncommutative Space}
\author{Guang-Hong Chen, 
and Yong-Shi Wu}
\address{Department of Physics, University 
of Utah\\ Salt Lake City, Utah  84112\\
\vspace{.5cm} 
{\tt ghchen@physics.utah.edu \\
     wu@physics.utah.edu}
}

\maketitle

{\tighten
\begin{abstract}
In this paper we demonstrate that coordinate 
noncommutativity at short distances can 
show up in critical phenomena through UV-IR 
mixing. In the symmetric phase of the 
Landau-Ginsburg model, noncommutativity is 
shown to give rise to a non-zero anomalous 
dimension at one loop, and to cause 
instability towards a new phase 
at large noncommutativity. In particular, 
in less than four dimensions, the one-loop 
critical exponent $\eta$ is non-vanishing 
at the Wilson-Fisher fixed point.
\end{abstract}
}
\newpage


\section{Introduction}
Space quantization is an old idea. Specific 
coordinate commutation relations were 
suggested more than half century 
ago\cite{Snyder,Yang}, in an attempt
to resolve the short-distance singularities 
in local quantum field theory. Recently, 
interests in field theories 
on a noncommutative space (or simply NCFT) 
have been revived\cite{connes}, particularly 
because of their emergence in string/M(atrix) 
theory\cite{matrix,witten}. Also NCFT is 
expected to be relevant to planar quantum 
Hall systems in a strong magnetic field, 
since the guiding-center coordinates for 
the cyclotron motion of a charge in the 
lowest Landau level are known not to commute
with each other. 

The simplest case we have encountered in 
these instances is a space with constant 
noncommutativity: 
\begin{equation}
 [x^\mu, x^\nu]=i\Theta^{\mu\nu} 
\qquad (\mu, \nu = 1,2,\cdots,d),
\label{noncom}
\end{equation}
where $\Theta^{\mu\nu}=-\Theta_{\mu\nu}$ 
are real parameters, of dimension length 
squared. Classical field theory in such a 
space can be realized as a deformation of 
the usual field theory in an ordinary 
(commutative) space, by changing the 
product of two fields to the Moyal star 
product\cite{moyal} defined by
\begin{equation}
\label{star}
(f*g)(x)=\exp (\frac{i}{2}\Theta^{\mu\nu}
\partial^{x}_{\mu} \partial^{y}_{\nu}) 
f(x)g(y)|_{y=x}.
\end{equation}
Note that the first term on the right side
gives the ordinary product, recovered in
the limit $\Theta_{\mu\nu}\to 0$. Also the 
commutatator (\ref{noncom}) is realized as 
$[x^\mu, x^\nu]= x^\mu * x^\nu - x^\nu * x^\mu$.

An interesting issue of great importance for
future applications of NCFT is whether spatial
noncommutativity at short distances could show 
up in the low energy effective theory, or in
the critical behavior at large distances. 
Conventional wisdom seems to point to a negative
answer: When the distance scale under consideration 
is much larger than the length scale given by the 
coordinate noncommutativity, the effects of the 
latter should be negligible. More concretely, this
argument goes as follows: The essential difference
between the star and ordinary product is the terms 
other than the first one in eq. (\ref{star}), 
which all involve derivatives of higher orders.
It is well-known in theory of critical phenomena
that the interactions containing derivatives of
higher orders are all irrelevant. So one would
expect noncommutativity effects to vanish at 
sufficiently low energies.
   
The other side of the coin is that the commutator 
(\ref{noncom}) leads to the uncertainty relation 
$\Delta x^\mu \Delta x^\nu \sim \Theta_{\mu\nu}$. 
It tells us that for a wave packet in this space, 
if we make its size in $x^\mu$-direction small, 
then its size in $x^\nu$-direction will become 
big. So the ultraviolet (UV) effects in 
$x^\mu$-direction are entangled with the infrared 
(IR) effects in $x^\nu$-direction. Conceptually 
we believe it is such UV-IR mixing\cite{seiberg}, 
implied by coordinate commutation relations 
(\ref{noncom}), that makes noncommutativity 
effects capable of showing up at large distances. 
This motivated us to study, using perturbative 
renormalization group techniques, the critical 
behavior of the noncommutative Landau-Ginsburg 
model (NCLGM). As reported below, though the 
upper critical dimension remains to be four, 
we have found a number of effects that exhibit 
the $\Theta$-dependence at or near criticality. 
This confirms the entanglement between the UV 
(at the noncommutative scale, $\sqrt{\Theta}$) 
and the IR (at the scale of the correlation 
length) effects in a space described by eq. 
(\ref{noncom}).

\section{Noncommutative Landau-Ginsburg Model}
For simplicity, we first 
consider the $d=4$ case, with only 
$\Theta^{12}=\Theta^{34}=\Theta\neq 0$.
The Euclidean action of the Landau-Ginsburg 
model with a real scalar field reads
\begin{equation}
  \label{action}
S=-\int d^{4}x[\frac{1}{2}(\partial_{\mu}\phi)^2
+\frac{1}{2}m^2\phi^{2}
+\frac{g}{4!}\phi*\phi*\phi*\phi].
\end{equation}
Here the star product is defined in eq. 
(\ref{star}). Since the star product of two fields 
differs from the ordinary one only by a total 
divergence, the quadratic terms, as well as the
$\phi$-propagator, in NCLGM remain the same as 
in the ordinary Landau-Ginsburg model (OLGM). 
Also for a uniform phase with $\phi =const.$, 
the quartic terms are the same as in OLGM. So 
assuming uniform phases, the phase structure at 
tree level is the same as in the ordinary case: 
The critical point is still given by $m^2=0$, 
separating the symmetric phase (with $m^2>0$ 
$\langle \phi \rangle =0$), and the symmetry 
broken phase (with  $m^2< 0$ and 
$\langle \phi \rangle = const \neq 0$).
In this paper we consider only approaching 
to criticality from the symmetric phase.

The star product in the quartic term is 
invariant under only cyclic permutations of 
the factors. Using path integral, the Feynman 
rule in momentum space for the $\phi^4$ 
vertex acquires an additional phase factor:
\begin{eqnarray}
\label{phase}
\exp [-(i/2)\sum_{i<j} k_i \wedge k_j],
\end{eqnarray}
where $k_i$ are the cyclically ordered momenta 
flowing into the vertex, and $k_i \wedge k_j 
\equiv k_{i,\mu}\Theta^{\mu\nu} k_{j,\nu}$.
To keep track of the ordering of the lines, we 
either allow lines crossing over each other, 
or introduce a double-line representation for 
the $\phi$-propagator as if it is a marix.
Either way, one can classify Feynman diagrams 
into planar and non-planar graphs\cite{filk}.
Alternatively, to simplify the symmetry 
factors, one can symmetrize the phase factor
(\ref{phase}), resulting in the momentum 
space form of the quartic interactions:
\begin{equation}
\label{s4}
S_4=-\frac{1}{4!}\int^{\Lambda}\phi(4)
\phi(3)\phi(2)\phi(1)u(4321),
\end{equation}
where $\int^{\Lambda}=
\int^{\Lambda}_0\biggl(\prod_{i}
\frac{d^{4}k_i}{(2\pi)^4}\biggr) 
(2\pi)^4 \delta^{(4)}(\sum_ik_i)$.
The function $u(4321)$ is given by
\begin{eqnarray}
\label{interaction}
u(4321)&=&\frac{g}{3}\biggl[
\cos(\frac{k_1\wedge k_2}{2})
\cos(\frac{k_3 \wedge k_4}{2})\\ \nonumber
&+& 
\cos(\frac{k_1\wedge k_3}{2})
\cos(\frac{k_2\wedge k_4}{2}) \\ \nonumber
&+& \cos(\frac{k_1\wedge k_4}{2})
\cos(\frac{k_2\wedge k_3}{2})\biggr].
\end{eqnarray}

It turned out\cite{filk} that a planar graph 
always gives the same contribution in OLGM 
(with $\Theta=0$) multiplied by a phase 
factor, that depends only the external momenta 
and their cyclic ordering. Thus, planar graphs 
share the same UV divergences as in OLGM, so 
we still need a cut-off $\Lambda$ in momentum 
space to regulate the NCFT. But the behavior 
of non-planar graphs is very different from 
their $\Theta=0$ counterpart: Rapid 
oscillations of internal momentum dependent 
phase factors in the integrand make 
non-planar graphs in NCFT less divergent than 
in the ordinary theory. For example, vertex 
corrections at one loop in ordinary $\phi^4$ 
theory are known to be logarithmically 
divergent; from this one guesses that 
non-planar vertex diagrams should be finite 
in NCLGM. Indeed, this can be confirmed by 
explicit calculations (see below).

Perturbative calculations in this NCFT have 
been done at one \cite{seiberg} and 
two\cite{arefeva} loops. It was found that 
for the inverse propagator, non-planar 
graphs make the two limits, the cut-off 
$\Lambda\to \infty$ and the external 
momentum $k\to 0$, not interchangeable. If 
one insists to have $\Lambda\to \infty$ 
first, then the propagator becomes singular
as $k\to 0$, seemingly obstructing a 
Wilsonian RG analysis\cite{seiberg}. We 
observe that this potential IR problem can 
be bypassed by using a shell integration 
in momentum space in RG analysis, as 
advocated by Shankar\cite{shankar} and 
Polchinski\cite{polchinski}.

\section{RG analysis in $d$ dimensions} 
According to ref. \onlinecite{shankar}, 
the RG transformation can be found by using 
(the cumulant expansion with) usual Feymann 
diagrams, with the loop integrals being 
over a thin shell $\Lambda/s\le k \le\Lambda$ 
in momentum space ($s>1$). This shell 
integration represents the elimination of 
the fast modes $\phi_f(k)$ with momentum $k$ 
within the above shell, resulting in an 
effective action for the slow modes $\phi_s(k)$ 
with $0\le k\le\Lambda/s$. Then we rescale the 
momentum $k \to sk$ and field variables, 
to make the effective action to be of the same 
form as the original one, but with a new set 
of coupling constants. The relations between 
the new and the old coupling constants define
the {\it RG transfomation} of the system. As 
usual, we will adopt dimensional regularization, 
in which we first finish tensor operations 
involving $\Theta_{\mu\nu}$ in four dimensions, 
then continue the loop integrals with scalar 
integrand to $d$ dimensions. 

Applying this procedure to the quadratic 
term $S_0= -\frac{1}{2} \int^{\Lambda}k^2 
\phi(-k)\phi(k)$ and requiring it be a 
fixed-point action, leads to the following
RG transformations 
\begin{equation}
\label{rgtrans}
k^{\prime}=sk, \hspace{1.0cm} \phi^{\prime}
(k^{\prime})=s^{-\frac{d+2}{2}}\phi_s (k).
\end{equation}
Similarly for the term $S_2=
-\frac{m^2}{2}\int^{\Lambda}\phi(-k)\phi(k)$,
the mode elimination by shell integration, 
together with RG transformation (\ref{rgtrans}), 
leads to the tree level RG transformation:
\begin{equation}
\label{rgmsquare}
m^{\prime 2}=s^2 m^2.
\end{equation}
So the quadratic $S_2$ is  relevant in 
the RG sense. So far there is no difference 
between NCLGM and OLGM.

To proceed, if one naively expand the cosine 
factors in the quartic term (\ref{interaction})
in powers in momenta, and apply the above 
RG transformation to each term, then at 
tree level this leads to the conclusion 
that all the $\Theta$ dependent terms
are irrelevant in any dimension. However, 
this expansion does not respect the star 
product structure, the intrinsic feature of 
any field theory on an NC space, that 
coherently organizes infinitely many higher 
order derivative terms. Though each of them
behaves like irrelevant, their coherent sum
may gives rise to non-trivial effects. 
Indeed, at least at one loop, it has been 
shown\cite{seiberg} that counterterms have 
the same star product structure. Thus, the 
operators allowed to appear in the Wilsonian 
effective action must be always of the the 
form of a star product with the same $\Theta$ 
parameter. The generic quartic terms are
always of the form of eq. ({\ref{interaction}), 
with the prefactor $g$ a function of momenta. 
Therefore, we should classify the terms in the 
square bracket in eq. (\ref{interaction}) as 
a {\it marginal} operator and its marginality 
is protected from quantum fluctuations in NCFT
intrinsically by geometry. The difference of 
our treatment from the RG analysis in the 
commutative case is that we apply the usual 
RG transformation only to the coefficient 
$g(k_1,k_2,k_3,k_4)$, with the star product 
structure intact, and define its relevance, 
irrelevance etc as usual. Thus at tree level 
$g^{\prime}=s^{4-d}g$.

At one-loop level, quartic terms  
give rise to a tadpole diagram. (See Fig.~1.)



Note that the loop momentum runs only over the 
shell $[\Lambda/s,\Lambda]$. The tadpole diagram 
contributes a correction to the quadratic term
\begin{equation}
S^{\prime}_{2}=-\frac{1}{2}
\int^{\frac{\Lambda}{s}}_{0}\phi(-k)
\phi(k)\Gamma_2(k),
\end{equation}
where $\Gamma_2(k)$ is given by (for $s=1+t$ 
very close to unity) 
\begin{eqnarray}
\label{selfenergy}
\Gamma_2(k)&=&\frac{g}{6}\int_{\frac{\Lambda}{s}}^{\Lambda}
\frac{d^dp}{(2\pi)^d}\frac{2+\cos k\wedge p}{p^2+m^2},
\nonumber \\
&=&\frac{g}{6} K_d \frac{\Lambda^{d}t}{\Lambda^2+m^2}
(3-\frac{1}{8}\Theta^2\Lambda^2k^2).
\end{eqnarray}
where $K_d=S_{d}/(2\pi)^{d}$ and 
$S_{d}$ is the surface area of 
a unit sphere in $d$ dimensions. In the 
last line, we have retained only terms
up to second order in momentum $k$, 
since higher-order terms in $k$ are 
irrelevant.

The first term in eq. (\ref{selfenergy}) 
is $k$-independent, modifying the RG 
transformation (\ref{rgmsquare}) to
the following one-loop 
$\Theta$-independent RG equation
\begin{equation}
\label{rgeq1}
\frac{dr}{dt}=2r+\frac{1}{2}u(1-r)K_d,
\end{equation}
with dimensionless $r\equiv m^2/\Lambda^{2}$ 
and $u\equiv g\Lambda^{d-4}$; $r \ll 1$ is assumed.

However, in eq. (\ref{selfenergy}) the 
second term comes from non-planar graphs. 
It is both $\Theta$- and $k$-dependent, 
and is marginal under the RG transformation 
(\ref{rgtrans}), since {\it it modifies 
the kinetic term $S_0$}. Using field theory 
terminology, it gives rise to {\it wave 
function renormalization} for the $\phi$-field, 
that explicitly {\it depends on noncommutativity}. 
This is our key observation in this paper. With 
this term, the Gaussian fixed point action 
$S_0$ is modified to
\begin{eqnarray}
\label{s0mod}
S^{\prime}_0&=&-\frac{1}{2}\biggl[1-\frac{u}
{48} K_d(\Theta\Lambda^2)^2 t \biggr]
\int^{\frac{\Lambda}{s}}_0 k^2\phi(-k)\phi(k)
\end{eqnarray}
Using $1+ \gamma t \approx s^{\gamma}$, and 
introducing a dimensionless $\theta=\Theta\Lambda^2$ 
for noncommutativity, we see that the $\phi$-field 
acquires an anomalous dimension, modifying 
eq. (\ref{rgtrans}) to 
\begin{equation}
\label{scalinglaw}
\phi^{\prime}(k^{\prime})=s^{-\frac{d+2-
\gamma(u,\theta)}{2}}\phi(k)
\end{equation}
with the one-loop $\gamma$ depending on 
noncommutativity:
\begin{equation}
\label{anomalous}
\gamma(u,\theta)=-\frac{1}{48}uK_d\theta^2.
\end{equation}
This is a novel result, because in OLGM the 
anomalous dimension vanishes at one loop. 
It is a consequence of the nontrivial UV/IR 
mixing in NCLGM. Also unusual is that the 
{\it negative} value of this anomalous 
dimension. It may significantly affect 
the stability of the symmetric phase. We 
will come back to this point later. 

To calculate one-loop corrections to the 
quartic interaction vertex, as in OLGM
\cite{wilson}, we set all the external 
momenta to zero. With this prescription, 
a direct calculation shows that the 
one-loop RG equation in NCLGM for the 
quartic coupling $u$ is the same as in OLGM; 
i.e.,
\begin{equation}
\label{rgu}
\frac{du}{dt}=(4-d)u-\frac{3}{2}u^2K_d.
\end{equation}
With RG equations (\ref{rgeq1}),
(\ref{rgu}) and the anomalous dimension 
(\ref{anomalous}), we can now proceed to 
take a closer look on the new physics 
due to noncommutativity.

\section{In $d\geq 4$ dimensions}
In NCLGM, the upper critical dimension 
remains to be $d=4$. When $d>4$, the 
quartic coupling $u$ is irrelevant. So 
it is the unique, trivial Gaussian fixed 
point, $u^{*}=0=r^{*}$, that controls the 
IR asymptotically free low-energy behavior, 
with the same sets of critical exponents 
as usual in mean field theory: $\nu=1/2$ 
and $\eta=\gamma(u^{*},\theta)=0$. 

However, this is not the whole story, when 
we consider approaching to the critical point
$r=0$. In fact, the modified scaling law 
(\ref{scalinglaw}) gives a two-point 
correlation function behaving like 
\begin{equation}
\label{correlation}
\langle\phi(x)\phi(0)\rangle\sim\frac{1}
{x^{d-2 +\gamma}}.
\end{equation}
Due to the minus sign in (\ref{anomalous}), 
for very large noncommutative parameter 
$\theta$, the above correlation function 
does not diverge, which signals an 
instability of the system. The critical 
value, $\theta_c$, is given by condition
\begin{equation}
\label{condition}
u\theta^2_c=\frac{48(d-2)}{K_d}.
\end{equation}
More precisely, the parameter space for 
NCLGM is now three-dimensional, described
by $(u,r,\theta)$. The condition 
(\ref{condition}) gives us a surface 
in the parameter space. To access the 
Gaussian fixed point, we have to fine-tune 
the parameter $\theta$ to make $\theta < 
\theta_c(u)$.
Of course, the more close to the fixed 
point, the less important is the condition 
(\ref{condition}), since $\theta_c$ is 
pushed to infinity when arriving at the 
Gaussian fixed point.

In the critical dimension $d=4$, the 
one-loop non-zero anomalous dimension 
(\ref{anomalous}) is expected to modify 
the logarithmic corrections to the scaling 
laws at criticality.

\section{In $d=4-\varepsilon$ dimensions}

If the dimension is slightly lower than four, 
the Gaussian fixed point becomes unstable 
in IR, and we have a new IR stable fixed 
point, the noncommutative counterpart of the 
Wilson-Fisher (NCWF) fixed point. Besides the 
noncommutativity parameter $\theta$, its 
position in $(r,u)$-space for small 
$\varepsilon\equiv 4-d$ is the same 
as in usual OLGM:
\begin{equation}
\label{wf}
u^{*}=\frac{16\pi^2}{3}\varepsilon,\hspace{1.0cm}
 r^{*}=-\frac{1}{6}\varepsilon.
\end{equation}
At this fixed point, the critical exponent 
$\nu$ is unchanged: 
$\nu=\frac{1}{2}+\frac{\varepsilon}{12}$, but the 
one-loop critical exponent $\eta$ becomes
non-vanishing: 
\begin{equation}
\label{wfeta}
\eta= \gamma (u^{*}, \theta)=
-\frac{\varepsilon\theta^2}{72}.
\end{equation}
This result is characteristic of the NCLGM. 
We would like to stress three important 
aspects of the critical exponent (\ref{wfeta}): 
(1) It starts at order of $\varepsilon$, while
in OLGM it starts at order of $\varepsilon^2$
from a higher-order calculation. (2) It is 
{\it negative}, while it is positive in OLGM. 
(3) It looks non-universal because of its 
dependence on the dimensionless noncommutativity 
parameter $\theta$. However, in the present 
case, the NCWF fixed point had better be viewed 
as a line of fixed points labelled by $\theta$. 
Since $\theta$ originates in the microscopic
sector of the system, its appearance in the 
macroscopic critical exponent is a genuine 
manifestation of {\it UV/IR mixing}, namely, 
the fingerprint of a "high-energy" parameter
in low-energy phenomena.

To see how the anomalous dimension 
(\ref{wfeta}) affects the stability of 
the NCWF fixed point, let us examine the
two-point correlation function 
\begin{equation}
\label{corrscaling}
\langle\phi(x)\phi(0)\rangle\sim\frac{1}
{x^{2-\varepsilon (1+\theta^2/72)}}.
\end{equation}
The criterion to maintain the stability 
of the NCWF fixed point is that the above
correlation function should be divergent 
for short distances. However, now we have
the additional parameter $\theta$ as a 
new knob to tune the system. If it is 
too large, the correlation function can 
become convergent. The critical value is 
given by
\begin{equation}
\label{criticaltheta}
\theta_c=12/\sqrt{\varepsilon},
\end{equation}
which depends only on $\varepsilon$. 
Therefore, if $\theta<\theta_c$, the NCWF 
fixed point is stable. On the contrary, 
for $\theta>\theta_c$ the NCWF fixed point 
will no longer be stable. This is 
reflected in the phase diagram, Fig. 2.

This situation is similar to previous RG 
analysis for one dimensional and three 
dimensional interacting fermion systems. 
There RG analysis could be used to show a 
similar instability for the Fermi liquid 
fixed point. To get the picture of the 
Luttinger liquid (in 1$d$) and BCS 
superconductivity (in 3$d$), one had to 
determine nonperturbatively the underlying 
physics for the new phase. Here to gain 
knowledge of the new phase for $\theta>\theta_c$, 
we also need extra efforts. But the study of 
the new phase in the large $\theta$ limit 
is beyond the scope of the present paper.

\section{Conclusions and Discussions}

With the symmetric phase in NCLGM as an 
example, we have discussed the general 
features of critical phenomena on a
noncommutative space. We have demonstrated 
that small spatial noncommutativity does 
not change the phase structure of the 
system, its upper critical dimension (four), 
nor the RG flow equations for $r$ and $u$,
the positions of fixed points in  
($r$,$u$)-space and their nature. So the 
critical behavior in $d>4$ is exactly 
the same as in the commutative case. On 
the other hand, through UV-IR mixing, 
noncommutativity gives rise to a non-zero 
anomalous dimension for the order parameter. 
In four dimensions it modifies logarithmic 
corrections to the scaling laws in the 
critical theory, while it changes the
exponent $\eta$ at the NCWF fixed point in 
less than four dimensions. In fact, the 
NCWF fixed points form a line of fixed 
points, where the action contains a star 
product of $\phi^4$, labeled by the 
noncommutativity parameter $\theta\equiv 
\Theta \Lambda^2$, whose inverse is the 
analog of the magnetic flux per plaquette 
in a lattice model.

Our perturbative RG analysis starts with 
the assumption that the symmetric phase 
of NCLGM is continuously connected to that 
of OLGM, with $\langle \phi \rangle = 0$. 
Our results are consistent with this 
assumption for small noncommutativity, 
while indicating instability towards a 
new phase for large $\theta$, consistent 
with the phase diagram conjectured in 
ref. \onlinecite{gubser}. 

We expect that noncommutativity should 
also show up in the critical exponents in
the symmetry-broken phase in the NCLGM. 
Moreover, our analysis is consistent with 
the proposition, made recently in refs. 
\onlinecite{gubser} and \onlinecite{italy}, 
that the NCLGM is perturbatively renormalizable.
Details of our computations and generalization 
to the complex scalar and non-relativistic 
cases will be published elsewhere \cite{chenwu}.   

{\it Acknowledgment} \qquad One of us (GHC) 
acknowledges stimulating discussions with M.P.A. 
Fisher on possible relevance of noncommutativity 
in low energy physics. This research was supported 
in part by the U.S. NSF under Grant No. PHY-9970701.


\newpage
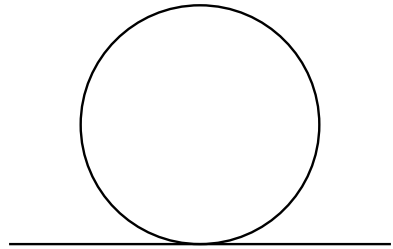
\begin{figure}
\caption{Tadpole diagram. }
\end{figure}
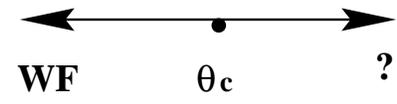
\begin{figure}
\caption{ Phase diagram along $\theta$ parameter. }
\end {figure}
\end{document}